\documentclass[12pt]{article}
\usepackage[a4paper, left=2cm, top=2cm, right=2cm, bottom=1.5cm, includefoot]{geometry}
\usepackage{amsmath}
\usepackage{amsthm}
\usepackage{amssymb}
\usepackage{bm}
\usepackage{tabularx}
\usepackage{arydshln}
\usepackage{graphicx}
\usepackage{subfigure}
\usepackage{color}
\usepackage[pdftex]{hyperref}
\usepackage{listings}
\usepackage{array}
\usepackage{cite}
\usepackage{rotating}
\usepackage[numbers,sort&compress]{natbib}

\hypersetup{%
  colorlinks=true,
  urlcolor=blue,
  citecolor=black
}

\title{Compact difference schemes for weakly-nonlinear parabolic and Schr\"odinger-type equations and systems}

\author{Vladimir A.~Gordin 
    \thanks{National Research University Higher School of Economics,
    Hydrometeorological Center of Russia, Moscow 123242, Russia. Email: vagordin@mail.ru (corresponding author)}
\and Evgenii A.~Tsymbalov \thanks{
    Skolkovo Institute for Science and Technology, Russia. Email: etsymbalov@gmail.com}}

\date{}

\begin{document}

\maketitle 

\begin{abstract}                                                                                          

The implicit compact finite-difference scheme is developed for evolutionary partial differential parabolic and Schr\"odinger-type equations and systems with a weak nonlinearity. To make any temporal step of the compact implicit scheme we need to solve a non-linear algebraic equations system. We use a simple explicit difference scheme for the first step followed by relaxation. Numerical experiments confirm the 4-th accuracy order of the algorithm.
The Richardson's extrapolation improves it up to the 6-th order.


\textbf{Key words:} compact high-order finite-difference scheme, parabolic equation, Schr\"odinger-type equation, 
weak non-linearity, Richardson extrapolation
\end{abstract}

\section{Introduction}

Weakly nonlinear partial differential equations and systems (parabolic and Schr\"odinger-like types) describe a wide spectrum of physical, physiological, ecological, genetic, etc phenomena   
\protect\cite{fisher1937wave, KPP1937, gross1961structure, novikov1984theory, murray1977lectures, bratus2010dynamical, dunbar1983travelling, patlak1953random, keller1971model, keller1971traveling, landau1965ld, abrikosov1957magnetic, fitzhugh1961impulses, schwan1969biological, nagumo1962active, pitaevskii1961vortex}.


We propose here an implicit compact finite-difference scheme for the approximation and numerical solution of such kind of models. The numerical scheme is effective for the mixed initial-boundary problems and provides the 4-th accuracy order. Earlier we considered such compact difference schemes for the  approximation of linear differential equations with  constant \cite{gord-14} and variable \cite{gt16a, gt16b, gord-10} coefficients. 


Here we develop an approach to the approximation of weakly non-linear partial differential equations or systems

\begin{equation}
\label{petro}
\partial_t \vec u =P(\partial_x)\vec u + F(x,\,\vec u),
\end{equation}
where $P(\partial_x)$ is a linear differential operator, and $\vec F$ is a given function, which is smooth with respect to both arguments. There are scalar and vector versions of the model, where the unknown function $\vec u$ is a scalar or vector function and, correspondingly, $P$ is a scalar or a matrix operator. For the scalar version we consider here the case 
\begin{equation}
\label{alpha}
P=\alpha \partial^2_x.
\end{equation}



The real part of the value $\alpha$ in the scalar version or the real part of the eigen-values in the matrix version are non-negative. Otherwise, the Cauchy problem for the (\ref{petro}) is incorrect.

We are based here on the coefficients of the compact schemes for the diffusion equation and Schr\"odinger equation with constant coefficients that were obtained in \cite{gord-14} (the compact schemes for such equations with variable coefficients were constructed in \cite{gordin2018compact}). However, such kind of compact scheme (with high approximation order) for other linear equations and systems, which are correct in the sense of I.G.~Petrovsky, see e.g. \cite{gelfand1967generalized}, may be found, and then they may be modified for the weak non-linear models.



The compact scheme is implicit, and for linear problems we need to inverse a matrix for every temporal step. To modify the approach for the case of non-linear equations and systems, it is necessary to solve at any temporal step a non-linear algebraic system. We use a simple explicit scheme to obtain a first guess of its solution. Then we do several relaxation iterations (see e.g. \cite{iserles2009first}, \cite{gordin2000mathematical}) to solve the original non-linear system with a good accuracy. 


Our numerical experiments confirm a high accuracy of such approach. The Richardson extrapolation method further improves the results and provides the 6-th accuracy order.


We consider here the Dirichlet boundary conditions only on both ends of the segment. However, the construction of such compact schemes can be weakly modified for other boundary conditions without the loss of the high accuracy order, see \cite{gord-10}, \cite{gordin2018compact}.

\section{Fisher -- Kolmogorov -- Petrovsky -- Piskunov (FKPP) model}
The Fisher -- Kolmogorov -- Petrovsky -- Piskunov equation

\begin{equation} \label{fkpp}
 \frac{\partial u}{\partial t}=D\frac{\partial^2 u}{\partial x^2} + \phi(u), \;D = const > 0,\;\phi \in {\bf C}^2
\end{equation}
describes a gene propagation \cite{fisher1937wave, KPP1937, bratus2010dynamical}. 
Here we take non-linear function: $\phi(u) = u(1-u)$ and the initial data $u(0, x) = u_0(x)$ with values on the segment $[0, 1]$, e.g.: $u_0(x) = cos(x), x \in [0, \pi/2]$. We will approximate non-linear partial differential equation (\protect\ref{fkpp}) on the grid $x_j=jh,\;h=\pi/2N,\;j=0,\ldots,N;\;t=n\tau,\;\tau$ is a temporal step, under Dirichlet conditions: 
\[
u^n_j \approx u(n\tau,jh),\; u^n_0=u^n_N=0.
\]


First, we approximate this equation by using the following implicit one-layer 4-th order compact difference scheme for the non-homogeneous linear diffusion equation with an arbitrary smooth forcing $f$:
\begin{equation} 
\label{diff_eq}
 \frac{\partial u}{\partial t}=D\frac{\partial^2 u}{\partial x^2} + f(t,\,x),\; D = const > 0,
\end{equation}
for which we have the following representation \cite{gord-14, gord-10}:
\[
a_0(u_{j-1}^{n+1} + u_{j+1}^{n+1}) + b_0 u_j^{n+1} + a_1(u_{j-1}^n + u_{j+1}^n) + b_1 u_j^n = 
\]
\begin{equation} 
\label{cds}
= p_0(f_{j-1}^{n+1} + f_{j+1}^{n+1}) + q_0 f_j^{n+1} + p_1(f_{j-1}^n + f_{j+1}^n) + q_1 f_j^n,\quad j=1,\ldots,N-1.
\end{equation}

Here we use the following coefficients: $a_0 = 2(6\nu-1); a_1 = -2(6\nu+1); b_0 = -4(6\nu+5); b_1 = 4(6\nu-5); 
p = p_0 = p_1 = -\tau; q = q_0 = q_1 = -10\tau$ to provide for scheme (\ref{cds}) the 4-th accuracy order. This scheme may be considered as a version of classic Crank-Nicholson scheme (see \cite{spotz1995high}) yet the full compact derivation is provided in \cite{gord-14}.


If we substitute the function $\phi(u)$ into equation (\protect\ref{diff_eq}) instead of $f(t,\,x)$, then we will get the following system of non-linear algebraic equations:
\begin{equation} 
\label{fkpp_system}
\begin{aligned}
    &a_0(u_{j-1}^{n+1} + u_{j+1}^{n+1}) + b_0 u_j^{n+1} = a_1(u_{j-1}^n + u_{j+1}^n) + b_1 u_j^n + 
    \\
    \\
    &+p(\phi(u_{j-1}^n) + \phi(u_{j+1}^n) + \phi(u_{j-1}^{n+1}) + \phi(u_{j+1}^{n+1})) + q (\phi(u_j^{n+1}) + \phi(u_j^n)),\quad j=1,\ldots,N-1.
\end{aligned}
\end{equation}

To make a temporal step: $n\tau \mapsto (n+1)\tau$  one needs to solve the system of non-linear equations with respect to unknown values $\{u_j^{n+1}\}_{j=1}^{N-1}$.


\subsection{The Explicit Euler scheme as the first guess and its subsequent improvement} \label{sect:fkpp-euler}


To solve numerically non-linear system (\protect\ref{fkpp_system}) we use $M=T/\tau$ times the following simplest algorithm of the approximate problem's integration on every temporal step:

\begin{itemize}
    \item Use the explicit Euler scheme to obtain a rough evaluation of unknown values $u_j^{n+1}$:
        \begin{equation*} 
            \hat{u}_j^{n+1} = u_j^n + \nu(u_{j-1}^n - 2 u_j^n + u_{j+1}^n) + \tau \phi(u_j^n).
        \end{equation*} 
    \item Apply relaxation to system (\protect\ref{fkpp_system}), which introduces the correction $\{ \delta_j \}_{j=1}^{N-1}$ to the ${\hat u}_j^{n+1}$ so that  $\widetilde{ u_j^{n+1}} = \hat{u}_j^{n+1} + \delta_j$.
    


    \item If for a given small $\delta>0$ the inequality $|\delta_j|\le \delta$ is fulfilled for all $j$, we finish this temporal step by putting:

\[
u_{j}^{n+1}=\widetilde{ u}_{j}^{n+1}.
\]

\item Otherwise we compute the following guess for the grid function $\left\{u^{n+1}_j\right\}_{j=1}^{N-1}$:

\[
\widetilde{\widetilde{u}}_{j}^{n+1}=\widetilde{u}_{j}^{n+1}+\delta_{j}
\]
and repeat the relaxation's step of the algorithm.

\item When the inequality $|\delta_j|\le \delta$ is fulfilled, we finish the iterations and use the last iteration's result $\left\{u^{n+1}_j \right\}_{j=1}^{N-1}$ as an approximate solution of  Eq.~(\ref{diff_eq}) in the moment $t=(n+1)\tau$.

\item Do next temporal step $(n+1)\tau \mapsto (n+2)\tau$ according to the Euler scheme.
    \end{itemize}

\subsection{Adams -- Bashforth explicit scheme as the first guess} \label{sect:fkpp-adams}


The Euler scheme has the first accuracy order only.
We can try to improve the first (explicit) part of the algorithm:

\begin{itemize}
    \item Use an explicit Euler scheme with half temporal step $\tau/2$ to obtain a basic estimate of the values
$u_j^{n+1/2}$:
        \begin{equation*} 
            \hat{u}_j^{n+1/2} = u_j^n + \nu(u_{j-1}^n - 2 u_j^n + u_{j+1}^n)/2 + \tau \phi(u_j^n)/2.
        \end{equation*}
    \item Then use an explicit central difference (leap-frog) scheme to obtain an evaluation of the values $u_j^{n+1}$:

        \begin{equation*} 
            \hat{u}_j^{n+1} = u_j^n + \nu
            (\hat{u}_{j-1}^{n+1/2} - 
            2 \hat{u}_j^{n+1/2} + 
            \hat{u}_{j+1}^{n+1/2}) + \tau \phi(\hat{u}_j^{n+1/2}).
        \end{equation*} 
        
 This two-step algorithm is named as the Adams -- Bashforth scheme; its  accuracy order with respect to time is equal to $2$.
        
    \item Then we apply relaxation to system (\protect\ref{fkpp_system}), and do several iterations until the values $\{ \delta_j \}_{j=1}^{N-1}$ will be small enough.


\end{itemize}

\subsection{Correction calculation}
\label{sect:correct}
General idea: we consider an error in the equation of Syst.~(\ref{fkpp_system}) that corresponds to a point with index $j$ and modify the value ${\hat u}^{n+1}_j$ to improve namely this equation.

We know the following values in the equation: ${u}^{n}_{j-1},\, {u}^{n}_{j},\,{u}^{n}_{j+1}$ and the first guess ${\hat u}^{n+1}_{j-1},\,{\hat u}^{n+1}_{j},\,{\hat u}^{n+1}_{j+1}$. We linearize in Eq.~(\ref{fkpp_system}) approximately\footnote{We assume here, that the first guess is accurate enough, i.e. the correction $\delta_j$ will be small, and we can neglect the high order terms of the Taylor expansion.} the non-linear function: 
\[\phi (u^{n+1}_{j})\approx \phi({\hat u}^{n+1}_{j})+\phi'({\hat u}^{n+1}_{j})\delta_j,\quad \delta_j= \left[u^{n+1}_{j} - {\hat u}^{n+1}_{j}\right],
\]
where ${}'$ (prime) denotes the derivative with respect to $u$:
\begin{equation} 
\label{fkpp_system1}
\begin{aligned}
    &a_0({\hat u}_{j-1}^{n+1} + {\hat u}_{j+1}^{n+1}) + b_0 \left[{\hat u}_j^{n+1} +\delta_j\right]\approx a_1(u_{j-1}^n + u_{j+1}^n) + b_1 u_j^n + 
    \\
    \\
    &+p(\phi(u_{j-1}^n) + \phi(u_{j+1}^n) + \phi({\hat u}_{j-1}^{n+1}) + \phi({\hat u}_{j+1}^{n+1})) + q\left[ (\phi({\hat u}_j^{n+1})+ \phi'({\hat u}^{n+1}_{j})\delta_j\right] + q\phi(u_j^n).
\end{aligned}
\end{equation}

We determine from relation (\ref{fkpp_system1}) the correction:
\begin{equation} 
\label{fkpp_system2}
\begin{aligned}
    &\delta_j = \left[b_0 - q\phi'({\hat u}^{n+1}_{j}) \right]^{-1}\left[-a_0({\hat u}_{j-1}^{n+1} + {\hat u}_{j+1}^{n+1})-b_0 {\hat u}_j^{n+1} +a_1(u_{j-1}^n + u_{j+1}^n) + b_1 u_j^n + \right.
    \\
    \\
    &\left.+p(\phi(u_{j-1}^n) + \phi(u_{j+1}^n) + \phi({\hat u}_{j-1}^{n+1}) + \phi({\hat u}_{j+1}^{n+1})) + q (\phi({\hat u}_j^{n+1}) + q\phi(u_j^n\right]).
\end{aligned}
\end{equation}

The value $b_0-q\phi'({\hat u}^{n+1}_{j})$ is strongly positive at sufficiently small temporal step $\tau$.

Then we improve the first guess of unknown value $ u_j^{n+1}$:
\[
{\hat u}_j^{n+1} \Rightarrow {\tilde u}_j^{n+1} ={\hat u}_j^{n+1} +\delta_j.
\]

Then we calculate the second modification $\left\{\tilde{\tilde u}^{n+1}_j\right\}_{j=1}^{N-1}$ by the first one $\left\{\tilde u^{n+1}_j\right\}_{j=1}^{N-1}$, etc. We stop the cycle, when the condition $|\delta_j|\le \delta$ is fulfilled.  

\textbf{Note 1.} One may try to modify the relaxation method by introducing into (\ref{fkpp_system2}) a relaxation parameter: $\frac{\omega}{b_0 -q\phi'({\hat u}^{n+1}_{j})}$ instead of $\frac{1}{b_0-q\phi'({\hat u}^{n+1}_{j})}$
to improve the iterations convergence
However, our numerical experiments show that the fine-tuning of $\omega$ parameter does not lead to an increase in efficiency of all the algorithms presented in this paper.

\subsection{Some details of the relaxation method} 
\label{sect:relax}
There are various orders to evaluate the corrections $\left\{\delta_j\right\}_{j=1}^{N-1}$. We list shortly some of them.

i) We calculate the values $\delta_j$ for every $j=1,\ldots,N-1$ and then add simultaneously these values to ${\hat u}^{n+1}_j$ and obtain a new grid function
${\tilde u}^{n+1}_j$.
Certainly, we do not obtain the exact solution of our Syst.~(\ref{fkpp_system}), because, when we modify a value ${\hat u}^{n+1}_j$, we introduce an error into the neighboring equations for $j-1$ and $j+1$. However, we believe (and there is a  proof of this statement, see e.g. \cite{iserles2009first, gordin2000mathematical}) that after such improvement of the first guess we decrease a norm of the error. When we repeat the algorithm several times, we obtain a solution of Syst.~(\ref{fkpp_system}) with a suitable accuracy.

ii) ``Chess modification'' of the version i). We calculate the the values $\delta_j$ for even indices $j$ only and add  them to the corresponding values ${\hat u}^{n+1}_j$. Then we use the modified even values $\left\{{\tilde u}^{n+1}_j\right\}$ to calculate the values $\delta_j$ for odd $j$. Then we repeat the modification for the even $j$ etc.

iii) We begin from the index $j=1$, calculate the value
$\delta_1$, and add it to the value ${\hat u}^{n+1}_1$. After this modification we go consequently to the index $j=2$, and use the modified value ${\tilde u}^{n+1}_1$ instead of the value ${\hat u}^{n+1}_1$. Then we go to the index $j=3$ etc. When we finish the process at $j=N-1$, we repeat the algorithm with obtained grid function. We begin again from the index $j=1$.

iv) ``Alternating direction modification'' of the version iii). When we finish at $j=N-1$ the algorithm iii), we begin second bypass 
\begin{equation}
    \label{rela}
    \left\{{\tilde u}^{n+1}_j\right\}\to \left\{{\tilde {\tilde u}}^{n+1}_j\right\}
\end{equation}
not from $j=1$, but from $N-1$ and will 
go by the grid, not from left to right but from right to left. And we alternate the bypass direction in algorithm~(\ref{rela}) after every pass.

v) We can go in~(\ref{rela}) from the ends of the segment to its center or from the center to the ends.

vi) We can divide the set $j=1,\ldots, N-1$ into several parts and realize these algorithms~(\ref{rela}) for every part separately. We can change the set's division after every cycle.

vii) We can apply the versions i-vi) in various combinations.

We have performed a series of numerical experiments to compare these versions and we had chosen the ``chess'' modification ii). Even a combination of such versions (e.g. version i) on the even iterations and version ii) on the odd ones) shows that the ``chess'' modification is better.

The effectiveness of the versions is different, but these differences are not dramatic. An optimal choice was essential for old computers, see e.g. \cite{gord1981}. But if we optimize the algorithm for parallel computations, the choice of an optimal version depends on the architecture of the particular computer, implementation and probably differs for every particular initial function $u_0(x)$ and non-linear function $\phi$.

\textbf{Note 2.} Instead of the relaxation, the Newton -- Raphson method will increase both the number of iterations required (even when paired with the double-sweep method) and computational time. Thus, relaxation (which is used result of an explicit finite-difference scheme's step as the first guess) is the most suitable for the compact finite-difference scheme implementation.

\section{Parabolic weakly-nonlinear system}

The described method could be generalized for solution of weak non-linear parabolic systems, that are traditionally used in biology, see e.g. \cite{nagumo1962active, bratus2010dynamical, murray1977lectures, dunbar1983travelling, patlak1953random, keller1971model, keller1971traveling, ataullakhanov2002new, ataullakhanov2007intricate}.
If the matrix $\alpha$ in (\ref{alpha}) has a simple structure, we can reduce such system to the following quasi-diagonal form:


\begin{equation} 
\label{diff_syst}
    \begin{cases} 
        \frac{\partial u}{\partial t}=D_1\frac{\partial^2 u}{\partial x^2} + \phi_1(u, w); 
        \\ 
        \frac{\partial w}{\partial t}=D_2\frac{\partial^2 w}{\partial x^2} + \phi_2(u, w),\end{cases} D_1, D_2 > 0.
\end{equation}

As one of the examples of a weak non-linear parabolic quasi-diagonal system, we consider here the FitzHugh -- Nagumo model of biological neuron.  Here, $u$ is a membrane voltage and $w$ is a recovery variable:


\begin{equation} 
\label{fhn}
 \phi_1(u, w) = \epsilon(w - \alpha u - \beta),  \;
 \phi_2(u, w) = -(u - \mu w + w^3),                \;
 \alpha, \beta, \epsilon > 0; \mu \in \mathbb{R}.
\end{equation}

We solve this system with the Dirichlet boundary conditions. We can obtain the first guess for the solution on the $(n+1)$th time step by using the Euler scheme. 



\begin{equation*}
    \begin{cases} 
        {\hat {\vec u}}^{n+1} = \vec u^n + \nu_1 M \vec u^n + \tau \vec \phi_1(\vec u^n, \vec w^n); 
        \\ 
        {\hat {\vec w}}^{n+1} = \vec w^n + \nu_2 M \vec w^n + \tau \vec \phi_2(\vec u^n, \vec w^n),
    \end{cases}
\end{equation*}
where $M$ is a tridiagonal matrix with $-2$ values on a main diagonal and $1$ values on side ones. 

The Euler scheme's accuracy order is equal to $1$. To improve the accuracy the Adams -- Bashforth scheme should be used like Subsect.~\ref{sect:fkpp-adams}.

We use the same coefficients as in~(\ref{cds}) for the compact approximation of both the  equations,  linearization~(\ref{fkpp_system1}), and correction~(\ref{fkpp_system2}) for Syst.~(\ref{fhn}).

Then we use the first guess $\{\hat{u}^{n+1}_j\}_{j=0}^N$ to solve the non-linear algebraic system, which obtained as a compact approximation of Syst.~\ref{diff_syst}. Then we use the relaxation method ii) for every equation.

\section{Nonlinear Schr\"odinger equation}
Nonlinear Schr\"odinger equation (NLSE) is one of the most famous non-linear partial differential equations, see e.g. \cite{novikov1984theory}, \cite{smirnov2013constructed}. It describes many physical phenomena, e.g. in plasma physics, in oceanology, and in non-linear optics. The equation can be interpreted as an infinite dimensional Hamiltonian system, and the system is fully integrable; there are explicit soliton-like solutions of NLSE, see e.g. \cite{landau1965ld}. The non-linearity\footnote{Eq.~(\ref{schr}) may be used as a mathematical model for expansion of the 2D monolayer of the strongly interacting superfluid Fermi gas in the 3D vacuum, if we modify  here the last term and write  $-\frac{5}{2}|\psi|^{10/3}\psi$ instead of $\beta|\psi|^2 \psi$.} in the equation is smooth in the real sense, but it is not analytical with respect to the unknown complex-valued function $\psi (t,\,x)$:


\begin{equation} \label{schr}
    i\frac{\partial \psi}{\partial t} + \frac{\partial^2 \psi}{\partial x^2} + \beta|\psi|^2 \psi = 0.
\end{equation}

It can be rewritten as the following system of two real PDEs, which is similar to system (\protect\ref{diff_syst}): 

\begin{equation*}
    u = Re(\psi), w = Im(\psi), \phi_1(u, w) = -\beta(w^3 + w u^2), \phi_2(u, w) = \beta(w^3 + u w^2).   
\end{equation*}

We will use in our numerical experiments the following tentative solution (it is a soliton) for Eq.~(\protect\ref{schr}):


\begin{equation}\label{soliton}
    \psi = \frac{\sqrt{2 \alpha \beta^{-1}}}{cosh \sqrt{\alpha} (x - U t)} exp(-0.5 i U x + i (r^2 - \alpha) t)),
\end{equation}
where $\alpha$ and $U$ are the soliton's parameters. 


We repeat here an aforementioned compact scheme approach for the NLSE: a simple explicit scheme is used to obtain a first guess on a temporal step and then the relaxation iterations. There is a significant difference with system (\protect\ref{diff_syst}): the spectrum of the operator (\protect\ref{alpha}) is imaginary here against real and negative spectrum of the operator in system (\protect\ref{diff_syst}). However, our numerical approach works for both systems.

This approach also may be developed similarly for the Ginzburg -- Landau equation \cite{landau1965ld, abrikosov1957magnetic} as well as the Gross -- Pitaevsky equation \cite{gross1961structure, pitaevskii1961vortex}.



\section{Numerical experiments}
In order to examine the properties of the general approach for solving the problems (\protect\ref{fkpp}, \protect\ref{diff_syst}, \protect\ref{schr}), we conducted a series of numerical experiments. Note that for the two first problems we do not have analytic solutions, so we compared obtained solutions with reference ones that were calculated using a very fine mesh. 




\subsection{Accuracy and approximation order}
\label{subse.accur}

We will use below for the errors' evaluations the Chebyshev norm
\[
\|u(x)\|_C = \max\limits_{x\in [0,L]} |u(x)|,\quad \mbox{and}\;\;\|u_j\|_C = \max\limits_{j=1,\ldots, N-1} |u_j|,
\]
but the similar evaluations were obtained in the $L^2$-norm.

We integrate evolutionary equations and systems and compare reference and approximate solutions of the mixed initial-boundary problem in a time moment $T$. We choose the moment using the following criteria: 
\begin{itemize}
\item the difference between solutions of the considered problem in the moments $t=0$ and $t=T$ is essential;

\item in the moment $t=T$ our solution is far from the stationary\footnote{Parabolic equations and systems are dissipative, and (under usual boundary conditions) their solutions tend to stationary ones.Therefore, the good  approximation of these solutions at large $t$ should not be used for finite-difference schemes estimation.} one yet.
\end{itemize}




\subsection{Richardson extrapolation}
\label{subse.rich}
We can also use Richardson's extrapolation technique to improve the order by calculating the solution on a finer grid. If we obtain a family of approximate solutions $u=u_{h}(t,\,x)$ at $t=T$ and $\tau =h^2|\nu^{*}|/\max\limits_j D_j$ \footnote{Here $\nu^{*} = max(D_1, \,D_2) \tau h^{-2}$ in case of systems.}, 
and use the representation
\begin{equation}
\label{Ri}
u_h (T,\,x)= u(T,\,x)+h^4 u_*(T,\,x) + o(h^4),
\end{equation}
then we can calculate twice $u_h$ at $h=h_*$ and at $h=h_*/2$. After that we substitute the approximation into (\protect\ref{Ri}), neglecting the terms $o(h^4)$ and obtain the following algebraic system for two functions $u$ and $u_*$:
\begin{equation*}
u_{h_*}(T,\,x)=u(T,\,x)+h^4_* u_*(T,\,x),\; u_{h_*/2}(T,\,x)=u(T,\,x)+h^4_* u_*(T,\,x)/16 \Rightarrow
\end{equation*}
\[
\Rightarrow u=u(T,x) \approx \left[ 16 u_{h_*/2}(T,x) - u_{h_*}(T,x)\right]/15.
\]

The stopping criteria for the relaxation procedure everywhere is $\delta < 10^{-12}$, except for the case of NLSE, where $\delta < 10^{-8}$ is used.

\subsubsection{FKPP}
\label{examp1}
Numerical experiments on the tentative solutions of FKPP equation (\protect\ref{fkpp}) at $u_0 =\cos^2 (x), \;\phi(u)= u(1-u)$ are represented for several time moments $T$ on the Fig.~1. They confirmed $4$th order of the compact scheme (\ref{cds}) for both Euler- and Adams -- Bashforth-based versions, see Table~\ref{tab:fkpp-order}.

\begin{figure}[h!]
\begin{center}
    \includegraphics[scale=.4]{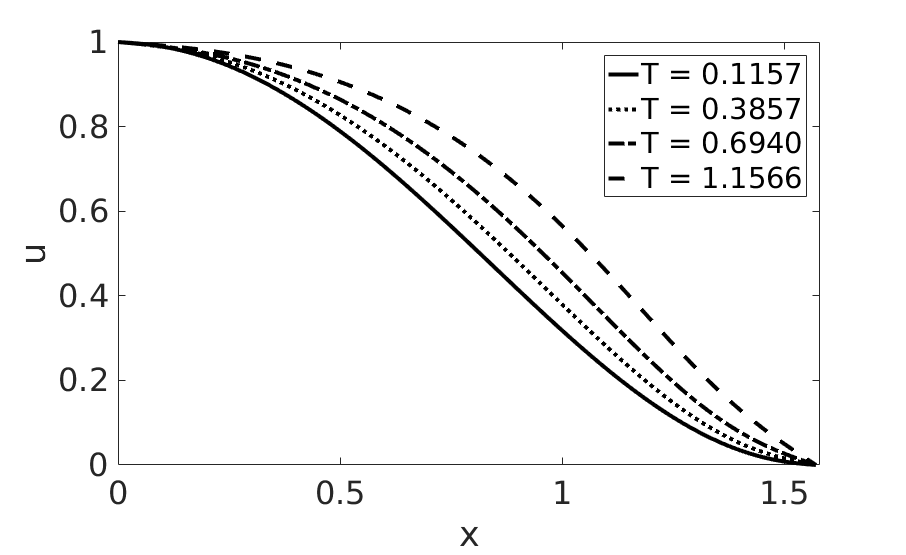}
    \caption{Solution of FKPP Eq.~(\protect\ref{fkpp}) for various integration times $T$, $D = 0.01,\; \nu = 0.3,\; N = 128,\; u_0 = cos^2(x),\; \phi(u) = u(1-u)$. We did not used a Richardson extrapolation here since the difference is not visible in illustrative context.} 
    \label{fig:fkpp-time}
\end{center}
\end{figure}
The accuracy does not depend much on the type of initial guess (Adams -- Bashford  or Euler), see Table~\protect\ref{tab:fkpp-order}. However, Adams -- Bashford scheme is more preferable specifically for the case of the large number $N$ of spatial grid knots, see Table~\ref{tab:fkpp-iters}.

However, for both first guess versions we obtain a fast convergence to an accurate solution, see the results of experiments with a fixed threshold $\delta$, see Table~ \protect\ref{tab:fkpp-iters}.



\begin{table}[h!]
\centering
\caption{Compact scheme for the FKPP equation's solution. We evaluate the error in {\bf C}-norm and error rate for the various nodes number $N$ and Courant parameter value $\nu$. Here we choose initial function $u_0 = cos^2(x)$, and parameters' values $D = 0.01,\; T = 4.63$. Initial guess type (Sect. \ref{sect:fkpp-euler} and \ref{sect:fkpp-adams}) does not affect the accuracy. We can see $4$th error rate. Here $\delta = 10^{-12}$.}
\label{tab:fkpp-order}
\begin{tabular}{|c|c|c|c|c|}
\hline
$\nu$ & \multicolumn{2}{c|}{0.1} & \multicolumn{2}{c|}{3.2} \\ \hline
$N$ & Error & Error rate & Error & Error rate \\ \hline
16 & 5.65-4 & - & 6.43-2 & - \\ \hline
32 & 3.23-5 & 4.12 & 2.21-3 & 4.86 \\ \hline
64 & 2.05-6 & 3.98 & 1.29-4 & 4.09 \\ \hline
128 & 1.29-7 & 3.99 & 8.10-6 & 4.00 \\ \hline
256 & 1.09-9 & 3.56 & 5.09-7 & 3.99 \\ \hline
\end{tabular}
\end{table}

\begin{table}[h!]
\centering
\caption{Compact scheme with a Richardson extrapolation (\ref{Ri}) for the FKPP equation's solution. We evaluate the error in {\bf C}-norm and error rate  for various values $\nu$. Here $u_0 = cos^2(x),\; D = 0.01,\; T = 4.63$. Initial guess type (Sect.~\ref{sect:fkpp-euler} and \ref{sect:fkpp-adams}) does not affect the accuracy.  A very high (6+) error rate is reached. The nodes number $N=64$ already reaches the accuracy limit (solution is compared with the one on the finer grid). The number $N=128$ is excessive. Here $\delta = 10^{-12}$.}
\label{tab:fkpp-richardson}
\begin{tabular}{|c|c|c|c|c|}
\hline
$\nu$ & \multicolumn{2}{c|}{0.8} & \multicolumn{2}{c|}{3.2} \\ \hline
$N$ & Error & Error rate & Error & Error rate \\ \hline
16 & 1.74-5 & - & 6.29-3 & - \\ \hline
32 & 1.63-7 & 6.73 & 9.09-6 & 9.43 \\ \hline
64 & 4.93-9 & 5.04 & 3.51-8 & 8.02 \\ \hline
128 & 4.99-9 & - & 4.93-9 & 2.82 \\ \hline
\end{tabular}
\end{table}

\subsubsection{FitzHugh -- Nagumo system}
\label{subse.FHNs}
\begin{figure}[h!]
\begin{center}
    \includegraphics[scale=.4]{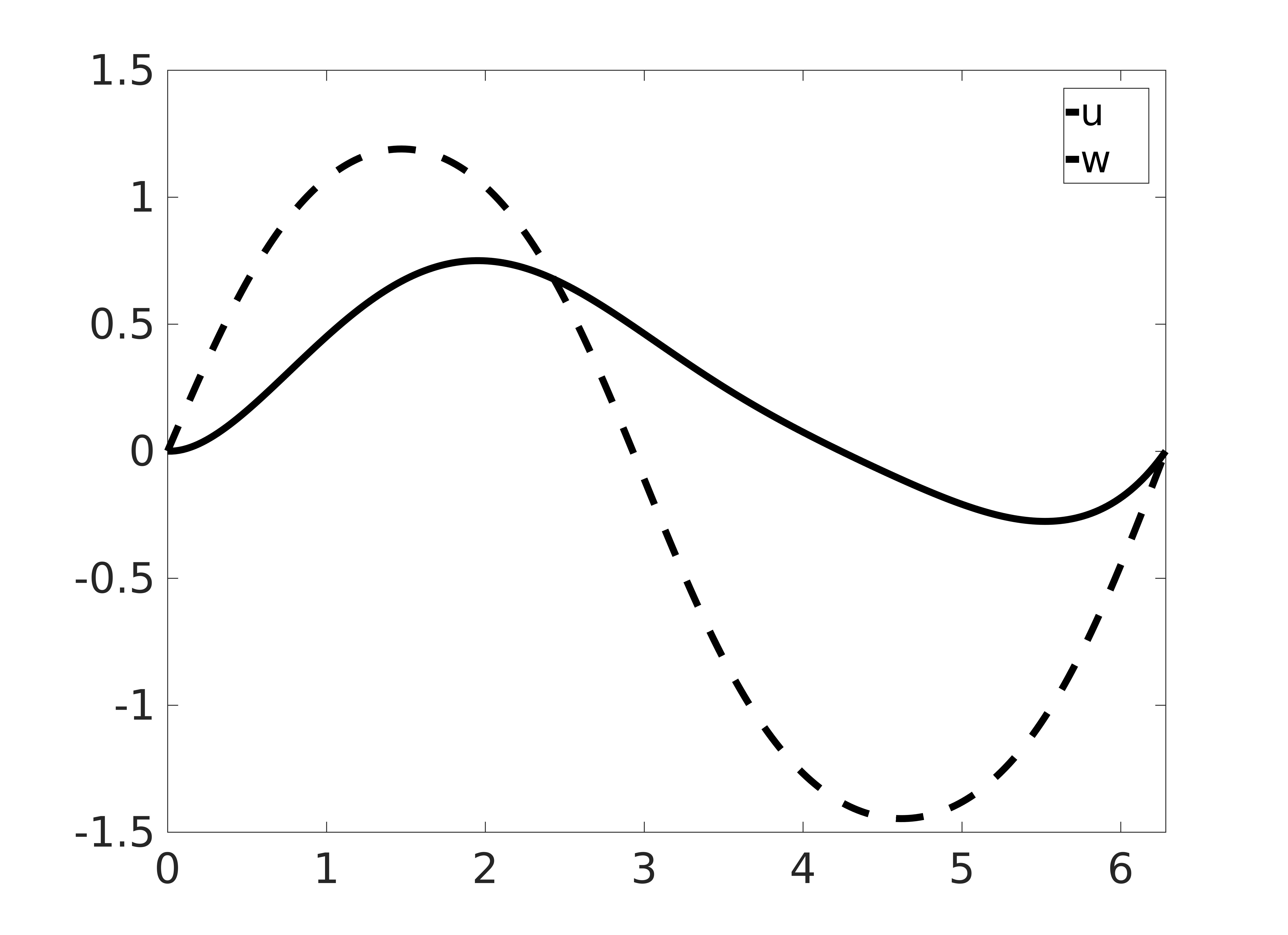}
    \caption{Solution of FHN Eq.~(\protect\ref{fhn}). $u_0 = \sin(x),\; w_0 = \sin^2(x),\; T = 0.2467, \epsilon = \alpha = \mu = 2, \beta = D_1 = D_2 = 1$. We did not use a Richardson extrapolation here since the difference is not visible in illustrative context.}
    \label{fig:fhn-solution}
\end{center}
\end{figure}

Numerical experiments on FitzHugh -- Nagumo system (\protect\ref{fhn}) show the 4th order of the Euler-based compact difference scheme (\protect\ref{cds}). We evaluated the order for both the components of solutions $u$ and $w$. See Fig.~\ref{fig:fhn-solution} for the solution we used in the numerical experiments. Adams -- Bashford initial guess is used, see Table~3.

\begin{table}[h!]
\centering
\caption{Compact scheme for the FHN system's solution. We evaluate the error in {\bf C}-norm and error rate for a various $\nu$. Here $u_0 = \sin(x),\; w_0 = \sin^2(x),\; T = 0.2467$. High $4$th accuracy order is reached. Here $\delta = 10^{-12}$.}
\label{tab:fhn-order}
\begin{tabular}{|c|c|c|c|c|c|c|}
\hline
$\nu$ & \multicolumn{3}{c|}{0.1} & \multicolumn{3}{c|}{0.4} \\ \hline
$N$ & Error for $u$ & Error for $w$ & Mean error rate & Error for $u$ & Error for $w$ & Mean error rate \\ \hline
8   & 4.25-3 & 9.79-3 & -    & -      & -      & -    \\ \hline
16  & 3.00-4 & 5.81-4 & 3.94 & 6.02-2 & 3.52-2 & -    \\ \hline
32  & 2.04-5 & 3.52-5 & 3.96 & 2.88-3 & 1.83-3 & 4.33 \\ \hline
64  & 1.28-6 & 2.19-6 & 4.01 & 1.77-4 & 1.13-4 & 4.02 \\ \hline
128 & 7.93-8 & 1.36-7 & 4.01 & 1.10-6 & 7.04-6 & 4.00 \\ \hline
256 & 4.42-9 & 7.84-9 & 4.14 & 6.89-7 & 4.41-7 & 4.00 \\ \hline
\end{tabular}
\end{table}

The Richardson extrapolation technique is also applicable here, and it improves the accuracy order up to the $6$th, see Table~4.

\begin{table}[h!]
\centering
\caption{Compact scheme for the FHN system's solution with Richardson extrapolation (\ref{Ri}). We evaluate the error in {\bf C}-norm and error rate  at various values $\nu$. Initial functions $u_0 = sin(x), w_0 = sin^2(x)$ and parameters' value are $\epsilon = \alpha = \mu = 2, \beta = D_1 = D_2 = 1, T = 0.2467$. The nodes numbers $N=64$ already provides the accuracy limit (solution is compared with the one on the finer grid). A very high (6+) error rate is reached. The nodes number $N=64$ already reaches the accuracy limit (solution is compared with the one on the finer grid). The number $N=128$ is excessive. Here $\delta = 10^{-12}$.}
\label{tab:fhn-richardson}
\begin{tabular}{|c|c|c|c|c|c|c|}
\hline
$\nu$ & \multicolumn{3}{c|}{0.1} & \multicolumn{3}{c|}{0.4} \\ \hline
$N$ & Error for $u$ & Error for $w$ & Mean error rate & Error for $u$ & Error for $w$ & Mean error rate \\ \hline
8   & 4.13-5 & 1.56-4 & -    & 1.00-3 & 5.48-4 & -    \\ \hline
16  & 3.47-7 & 1.23-6 & 6.94 & 2.52-6 & 1.61-6 & 9.49 \\ \hline
32  & 4.77-9 & 1.91-8 & 6.09 & 1.03-8 & 2.07-8 & 7.11 \\ \hline
64  & 1.26-9 & 1.31-9 & 2.90 & 1.22-9 & 1.30-9 & 3.53 \\ \hline
128 & 1.28-9 & 1.20-9 & -    & 1.22-9 & 1.23-9 & - \\ \hline
\end{tabular}
\end{table}

\subsubsection{Nonlinear Schr\"odinger equation}

Compact difference scheme (\protect\ref{cds}) may be used to solve non-linear Schr\"odinger equation (\protect\ref{schr}), too. Even when it is rewritten as the system of PDEs, this problem is completely different from (\protect\ref{diff_syst}), because the spectrum of the problem is imaginary in comparison to the  spectrum of parabolic system. We substitute $\psi=u+iv$ into  Eq.~(\protect\ref{schr}) and rewrite it as a PDE system for real functions $u$ and $v$: one should split the values in formula~(\ref{fkpp_system2}) into their real and imaginary parts and obtain finite-difference equations for the real functions $u$ and $v$, correspondingly.

\begin{figure}[h!]
\hspace{-1.5cm}
    \includegraphics[scale=.4]{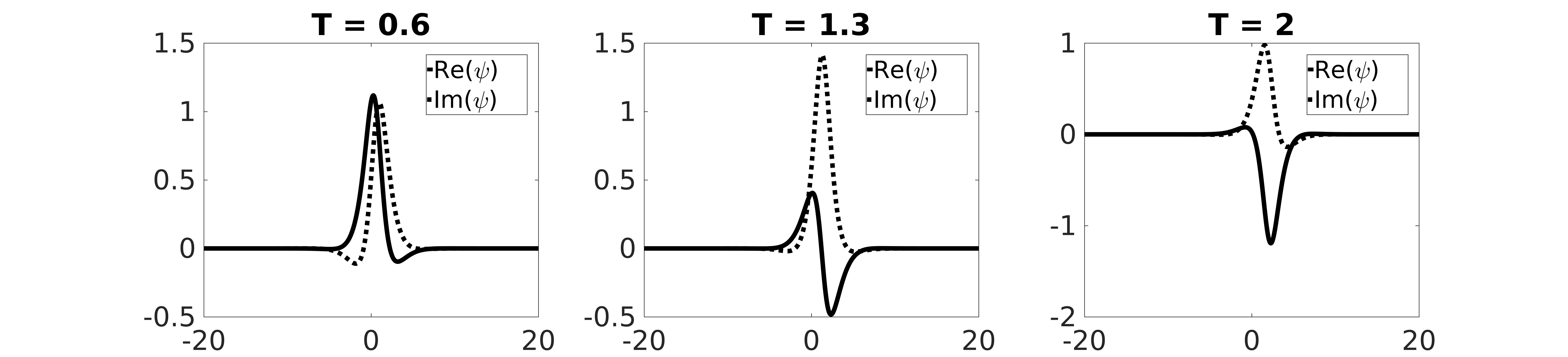}
    \caption{Real and imaginary parts of soliton-like solution of NLS Eq.~(\protect\ref{schr}) for different integration times $T$. Here $\alpha = \beta = U = 1$. }
    \label{fig:schr-solution}
\end{figure}

The solution for our numerical experiments see Fig.~\ref{fig:schr-solution}. Numerical experiments confirm the $4$th accuracy order of the compact scheme for NLSE equation, see Table~5. The Richardson extrapolation technique is also applicable here, and it improves the accuracy order up to the $6$th, see Table~6. Adams -- Bashford initial guess is used. The solutions we used are based on the Dirichlet boundary condition, where the actual values on the boundaries depend on integration time and are taken from the analytic solution.

\begin{table}[h!]
\centering
\caption{Compact scheme for the NLS equation's solution. We evaluate the error in {\bf C}-norm and error rate for various values $\nu$. Here $T = 1.25, \alpha = \beta = U = 1$. A high error rate of $4$ is reached. Here $\delta = 10^{-8}$.}
\label{tab:schr-order}
\begin{tabular}{|c|c|c|c|c|c|c|c|c|}
\hline
$\nu$ & \multicolumn{2}{c|}{0.05} & \multicolumn{2}{c|}{0.1} & 
    \multicolumn{2}{c|}{0.2} & \multicolumn{2}{c|}{0.4} \\ \hline
$N$ & Error  & Rate & Error  & Rate & Error  & Rate & Error  & Rate \\ \hline
32  & 4.38-1 & -    & 4.41-1 & -    & 4.52-1 & -    & -      & - \\ \hline
64  & 3.45-2 & 3.67 & 3.41-2 & 3.69 & 3.25-2 & 3.80 & 2.47-2 & - \\ \hline
128 & 1.61-3 & 4.42 & 1.59-3 & 4.42 & 1.49-3 & 4.45 & 1.19-3 & 4.38 \\ \hline
256 & 9.64-5 & 4.06 & 9.61-5 & 4.05 & 9.01-5 & 4.05 & 7.28-5 & 4.03 \\ \hline
512 & 8.68-6 & 3.47 & 4.73-6 & 4.35 & 4.30-6 & 4.39 & 4.52-6 & 4.01 \\ \hline
\end{tabular}
\end{table}

\begin{table}[h!]
\centering
\caption{Compact scheme for the NLS equation's solution with Richardson extrapolation (\ref{Ri}). We evaluate the error in {\bf C}-norm and error rate for various value $\nu$. $T = 1.25, \alpha = \beta = U = 1$. A very high (6+) error rate is reached. Here $\delta = 10^{-8}$.}
\label{tab:schr-richardson}
\begin{tabular}{|c|c|c|c|c|c|c|}
\hline
 & $\nu = 0.05$ & $\nu = 0.05$ & $\nu = 0.1$ & $\nu = 0.1$ & $\nu = 0.2$ & $\nu = 0.2$ \\ \hline
$N$ & Error  & Rate & Error  & Rate & Error  & Rate   \\ \hline
32  & 4.72-2 & -    & 4.70-2 & -    & 4.61-2 & -      \\ \hline
64  & 5.81-4 & 6.34 & 5.80-4 & 6.34 & 5.76-4 & 6.32   \\ \hline
128 & 4.78-6 & 6.92 & 4.08-6 & 7.15 & 5.19-6 & 6.79   \\ \hline
\end{tabular}
\end{table}


\subsection{Stability and Efficiency}
\label{subse.SE}



A criterion of stability is very important for various finite-difference schemes usage and comparison. Usually the criterion bounds its dimensionless Courant number $\nu=D\tau h^{-2}$. Our compact scheme is unconditionally stable for linear diffusion equation \cite{gord-14}. During our numerical experiments, we did not experience stability problems, even for large values of the Courant parameter $\nu \approx 50$.



We also conducted  many numerical experiments to find  the most efficient settings for the desired error, see Appendix \ref{appendix:eff} for detailed results.

\section{Summary and discussion}

We developed an approach for the high-order approximation of weak non-linear PDEs and systems. The approach was tested on the non-linear Schr\"odinger equation, the Fisher -- Kolmogorov -- Petrovsky -- Piskunov equation and the FitzHugh -- Nagumo model. Numerical experiments confirmed the $4$th order of the approach. We also show that our algorithm may be combined with the Richardson extrapolation technique, further improving the order up to the $6$th.

During the numerical testing of our approach, we did not experience stability issues, i.e. it seems to be unconditionally stable. 

The recommendation on efficient use of our algorithm is also included into the Appendices.

We have considered here the Dirichlet boundary conditions. However, the compact schemes may be applied for approximation under other boundary conditions. In such problems the function $f$ and its derivatives must be included into the approximate finite-difference boundary conditions to avoid loss of the approximation order, see \cite{gord-10}, \cite{gordin2018compact}.

\vspace{20pt}

The article was prepared within the framework of the Academic Fund Program at the National Research University --- Higher School of Economics (HSE) in 2016--2017 (grant No. 16-05-0069)  and by the Russian Academic Excellence Project "5-100".

\pagebreak
\appendix

\section{Appendix: The relaxation iterations number required for a good iterations' convergence}
\label{appendix:iters}
Initial AB approximation requires two times more computations than the Euler one. However, we would like to note that, on each temporal step, the initial approximation (Euler or AB) is much more lightweight than the relaxation iteration in terms of computational complexity. We compare here results of numerical experiments for small (in Table~7 $\nu=0.1$) and large (in Table~7 $\nu=3.2$). Then the  number of the temporal steps $T/\tau$ in the second variant is $32$ times smaller than in the first one.

\begin{table}[h!]
\centering
\caption{The average number of the relaxation iterations required for convergence for various values of Courant parameter $\nu$, initial approximations (explicit step: Euler or AB), and stopping criteria ($\delta$) for FKPP equation (\ref{fkpp}). $u_0 = \cos^2(x),\; D = 0.01,\; T = 4.63$. In most of the cases, AB scheme demonstrates reduced number of relaxation iterations compared to Euler scheme. No Richardson extrapolation was used here.}
\label{tab:fkpp-iters}
\begin{tabular}{|c|c|c|c|c|c|c|c|c|}
\hline
$\nu$ & \multicolumn{4}{c|}{0.1} & \multicolumn{4}{c|}{3.2} \\ \hline
$\delta$  & \multicolumn{2}{c|}{$10^{-6}$} & \multicolumn{2}{c|}{$10^{-12}$} &  \multicolumn{2}{c|}{$10^{-6}$} & \multicolumn{2}{c|}{$10^{-12}$} \\ \hline
$N$ & Euler & AB & Euler & AB & Euler & AB & Euler & AB \\ \hline
16 & 3 & 3 & 6 & 5 & 14 & 13 & 27 & 27 \\ \hline
32 & 3 & 2.4 & 5 & 5 & 17 & 13 & 39 & 35 \\ \hline
64 & 2 & 1.6 & 5 & 4.6 & 13 & 9 & 37 & 31 \\ \hline
128 & 1 & 1 & 4.2 & 4.1 & 9 & 7 & 32 & 27 \\ \hline
256 & 1 & 1 & 4 & 3.7 & 9 & 7 & 28 & 22 \\ \hline
\end{tabular}
\end{table}

We can conclude from Table~7 that, in the case of FKPP integration, there is a weak preference of the A-B scheme in comparison with the Euler one. The iterations' number increases slowly, when the constant $\delta$ decreases. The dependence of the number on the grid points number is non-monotonic.

The preferable values of $N, \nu, \delta$ in terms of computational time are listed in Table \ref{tab:fkpp-eff}. Large values of $\nu$ increase number of relaxation iterations for convergence and thus are not optimal.

\begin{table}[h!]
\centering
\caption{Average number of relaxation iterations required for convergence for various values of $\nu$, and stopping criteria for FHN equation (\ref{fhn}).  $u_0 = sin(x), w_0 = sin^2(x), \epsilon = \alpha = \mu = 2, \beta = D_1 = D_2 = 1, T = 0.2467$. Large values of $\tau$ and $\nu$ require more relaxation iterations. No Richardson extrapolation was used here.}
\label{tab:fhn-iters}
\begin{tabular}{|c|c|c|c|c|c|c|}
\hline
$\nu$ & \multicolumn{3}{c|}{0.1} & \multicolumn{3}{c|}{1.6} \\ \hline
$N \backslash \delta$ & $10^{-2}$ & $10^{-6}$ & $10^{-12}$ & $10^{-2}$ & $10^{-6}$ & $10^{-12}$ \\ \hline
8  & 3 & 7 & 12 & - & - & - \\ \hline
16 & 3 & 6 & 12 & 15 & 31 & 53 \\ \hline
32 & 3 & 6 & 12 & 15 & 32 & 57 \\ \hline
64 & 3 & 6 & 12 & 15 & 32 & 58 \\ \hline
128 & 3 & 6 & 12 & 14 & 32 & 58 \\ \hline
256 & 3 & 6 & 12 & 8 & 32 & 58 \\ \hline
\end{tabular}
\end{table}

Solution of the FHN problem requires more relaxation iterations than the solution of FKPP problem. The dependence on $N$ is very weak, and dependence on the parameter $\delta$ is essential.

\begin{table}[h!]
\centering
\caption{Average number of relaxation iterations required for convergence for a different $\nu$ for NLSE equation (\ref{schr}). Stopping criteria $\delta < 10^{-8}$. Large values of $\nu$ require more relaxation iterations. No Richardson extrapolation was used here.}
\label{tab:schrod-iters}
\begin{tabular}{|c|c|c|c|c|}
\hline
$N \backslash \nu$ & 0.05 & 0.1 & 0.2 & 0.4 \\ \hline
32 & 11 & 13 & 24 & 14 \\ \hline
64 & 10 & 12 & 19 & 460 \\ \hline
128 & 10 & 10 & 12 &  166 \\ \hline
256 & 9 & 10 & 10 &  115 \\ \hline
512 & 8 & 9 & 9 & 148 \\ \hline
\end{tabular}
\end{table}

\pagebreak
\section{Appendix: numerical results on the error as a function of relaxation stopping criteria.}
\label{appendix:relax-stop}

\begin{sidewaysfigure}[ht!]
    \includegraphics[width = \textwidth]{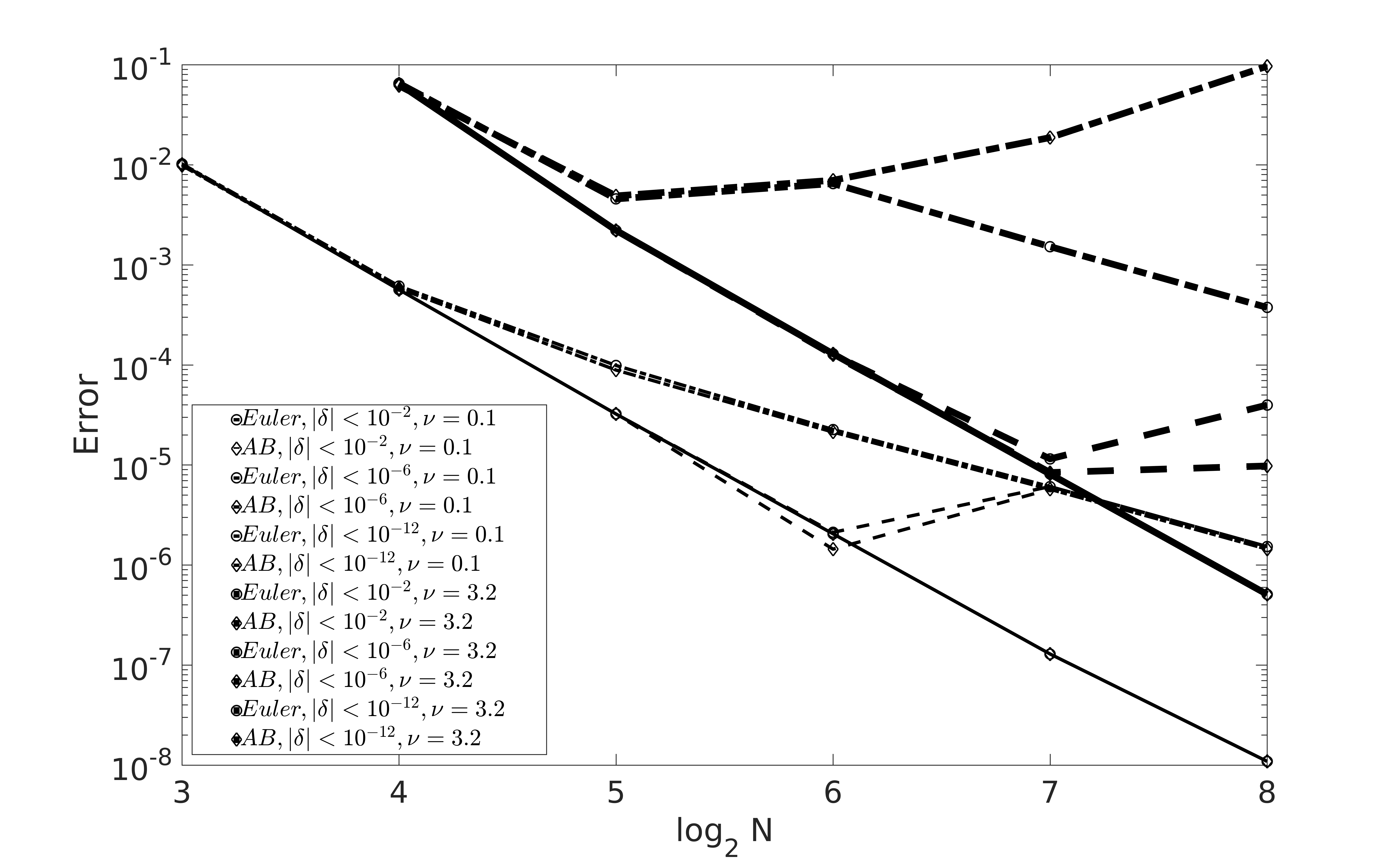}
    \caption{Error's norm as a function of the nodes number $N$, first guess, $\nu$, and stopping criteria for FKPP Eq.~(\protect\ref{fkpp}), $D = 0.01, u_0 = \cos^2(x), \phi(u) = u(1-u)$. Small $\nu$, AB initial guess type and small values for the stopping criteria result on the smallest error norm. Large $\delta$ may lead to convergence issues.No Richardson extrapolation was used here.} 
    \label{fig:fkpp-long-exper}
\end{sidewaysfigure}

The error of the final solution, i.e. the norm of the difference at the time moment $t=T$ between our solution and reference solution. It is the quality criterion of the scheme. We can decrease this difference, if we apply more expensive algorithms.  That is why, we should show the error for various values of the finite-difference scheme's parameters. See Fig. \ref{fig:fkpp-long-exper} for details.

\section{Appendix: Efficiency experiments.}
\label{appendix:eff}

We conducted a series of numerical experiments with various parameters that affects the calculation flow ($N$, $\nu$, stopping criteria $\delta$ for relaxation, denoted as "regime details") to find the most efficient (in terms of computational time) way to achieve the solution with the desirable accuracy. Our experiments demonstrate that one may increase the $N$ and $\delta$ to increase the accuracy in the efficient way.

\begin{table}[h!]
\centering
\caption{Optimal computational parameters for the compact finite-difference scheme approximation of FKPP Eq.~(\ref{fkpp}). Regime details: Courant parameter $\nu$, the  nodes' number $N$, initial approximations (explicit step: Euler or AB), and stopping criteria ($\delta$) in the least time-consuming case that reaches the desired error. Initial function $u_0 = \cos^2(x)$, the parameters value $D = 0.01,\; T = 4.63$. The most effective way to decrease the expected error is to increase $N$ and $\delta$. Large values of $\nu$ increase number of relaxation iterations for convergence and thus are not optimal. No Richardson extrapolation was used here.}
\label{tab:fkpp-eff}
\begin{tabular}{|c|c|c|c|c|c|c|}
\hline
Desired & \multicolumn{4}{c|}{Regime details} & \multicolumn{2}{c|}{Performance details} \\  \hline
Error & Guess & $N$ & $\delta$ & $\nu$ & Time (s) & Average \# of iterations \\ \hline
$10^{-2}$ & Euler & 16 & $10^{-2}$ & 0.8 & - & 2 \\ \hline
$10^{-3}$ & Euler & 16 & $10^{-2}$ & 0.4 & 6.3-5 & 1 \\ \hline
$10^{-4}$ & AB & 32 & $10^{-4}$ & 0.4 & 4.6-4 & 1.5 \\ \hline
$10^{-5}$ & AB & 64 & $10^{-6}$ & 0.8 & 2.0-3 & 3.3 \\ \hline
$10^{-6}$ & AB & 128 & $10^{-6}$ & 0.8 & 9.3-3 & 1.82 \\ \hline
$10^{-7}$ & AB & 256 & $10^{-8}$ & 0.8 & 1.0-1 & 2.95 \\ \hline
$10^{-8}$ & AB & 256 & $10^{-10}$ & 0.025 & 3.0 & 2.71 \\ \hline
\end{tabular}
\end{table}

\begin{table}[h!]
\centering
\caption{Optimal computational parameters value for the compact finite-difference scheme approximation of FHN equation (\ref{fhn}). Regime details: Courant parameter $\nu$, the number of grid nodes $N$, initial approximations (explicit step: Euler or AB), and stopping criteria ($\delta$) in the least time-consuming case that reaches the desired error. Initial functions are $u_0 = sin(x),\: w_0 = sin^2(x)$, the parameters value $\epsilon = \alpha = \mu = 2, \:\beta = D_1 = D_2 = 1, \:T = 0.2467$. The most effective way to decrease the expected error is to increase $N$ and $\delta$. Large values of $\nu$ increase number of relaxation iterations for convergence and thus are not optimal. No Richardson extrapolation was used here.}
\label{tab:fhn-eff}
\begin{tabular}{|c|c|c|c|c|c|}
\hline
Desired & \multicolumn{3}{c|}{Regime details} & \multicolumn{2}{c|}{Performance details} \\  \hline
Error & $N$ & $\delta$ & $\nu$ & Time (s) & Average \# of iterations \\ \hline
$10^{-2}$ & 8 & $10^{-2}$ & 0.1 & 4.6-4 & 3 \\ \hline
$10^{-3}$ & 16 & $10^{-2}$ & 0.2 & 8.8-4 & 3 \\ \hline
$10^{-4}$ & 32 & $10^{-4}$ & 0.2 & 4.1-3 & 4 \\ \hline
$10^{-5}$ & 64 & $10^{-6}$ & 0.2 & 2.2-2 & 5 \\ \hline
$10^{-6}$ & 128 & $10^{-8}$ & 0.4 & 1.7-1 & 13 \\ \hline
$10^{-7}$ & 256 & $10^{-10}$ & 0.4 & 1.61 & 16 \\ \hline
$10^{-8}$ & 256 & $10^{-10}$ & 0.1 & 5.27 & 10 \\ \hline
\end{tabular}
\end{table}

\bibliographystyle{ieeetr}
\bibliography{references} 

\end{document}